\documentclass[journal,twoside]{IEEEtran}
\usepackage{cite}
\usepackage{amsmath,amssymb,amsfonts}
\usepackage{algorithmic}
\usepackage{graphicx}
\usepackage{textcomp}
\usepackage{xcolor}
\usepackage{multirow}

\usepackage{booktabs}

\def\BibTeX{{\rm B\kern-.05em{\sc i\kern-.025em b}\kern-.08em
    T\kern-.1667em\lower.7ex\hbox{E}\kern-.125emX}}
\begin{document}

\title{
Stochastic Sparse Learning with Momentum Adaptation for Imprecise Memristor Networks
\thanks{Y. Wang and S. Wu contribute equally to this work.
Corresponding to: lpshi@mail.tsinghua.edu.cn}
}

\author{
Yaoyuan Wang$^{*}$, Shuang Wu$^{*}$, Ziyang Zhang, Lei Tian, Luping Shi$^{\dag}$  \\
Department of Precision Instrument, Center for Brain Inspired Computing Research\\
Beijing Innovation Center for Future Chip, Tsinghua University \\
}

\maketitle

\begin{abstract}
Memristor based neural networks have great potentials in on-chip neuromorphic computing systems due to the fast computation and low-energy consumption. However, the imprecise properties of existing memristor devices generally result in catastrophic failures for the network in-situ training, which significantly impedes their engineering applications. In this work, we design a novel learning scheme that integrates stochastic sparse updating with momentum adaption (SSM) to efficiently train the imprecise memristor networks with high classification accuracy. The SSM scheme consists of: (1) a stochastic and discrete learning method to make weight updates sparse; (2) a momentum based gradient algorithm to eliminate training noises and distill robust updates; (3) a network re-initialization method to mitigate the device-to-device variation; (4) an update compensation strategy to further stabilize the weight programming process. With the SSM scheme, experiments show that the classification accuracy on multilayer perceptron (MLP) and convolutional neural network (CNN) improves from 26.12\% to 90.07\% and from 65.98\% to 92.38\%, respectively. Meanwhile, the total numbers of weight updating pulses decrease 90\% and 40\% in MLP and CNN, respectively, and the convergence rates are both 3$\times$ faster. The SSM scheme provides a high-accuracy, low-power, and fast-convergence solution for the in-situ training of imprecise memristor networks, which is crucial to future neuromorphic intelligence systems.
\end{abstract}

\begin{IEEEkeywords}
memristors, neural networks, crossbar, momentum, neuromorphic
\end{IEEEkeywords}

\section{Introduction}
Recently, the field of neuromorphic computing has witnessed substantial advances in artificial intelligence applications achieved by deep neural networks (DNNs), such as image classification \cite{krizhevsky2012imagenet}, speech recognition \cite{hinton2012deep} and game playing \cite{silver2017mastering}. DNNs gain powerful generalization capabilities by their massive tunable weights, which are such data-intensive that the computation efficiency is usually limited by the memory access. Some DNN accelerators \cite{chen2017eyeriss} and neuromorphic chips \cite{merolla2014million,shi2015development} are designed with application-specific integrated circuit (ASIC) to alleviate the memory access bottleneck to speed up the computation. However, the synaptic weights are still stored in random access memories (RAM) rather than directly encoded by the states of emerging analogue devices. Such analogue device assembled systems are believed to achieve significant speedup and power reduction for on-chip deployments of DNNs \cite{yao2017face}.

\begin{figure}[bp]
\centering
\includegraphics[width=0.4\textwidth]{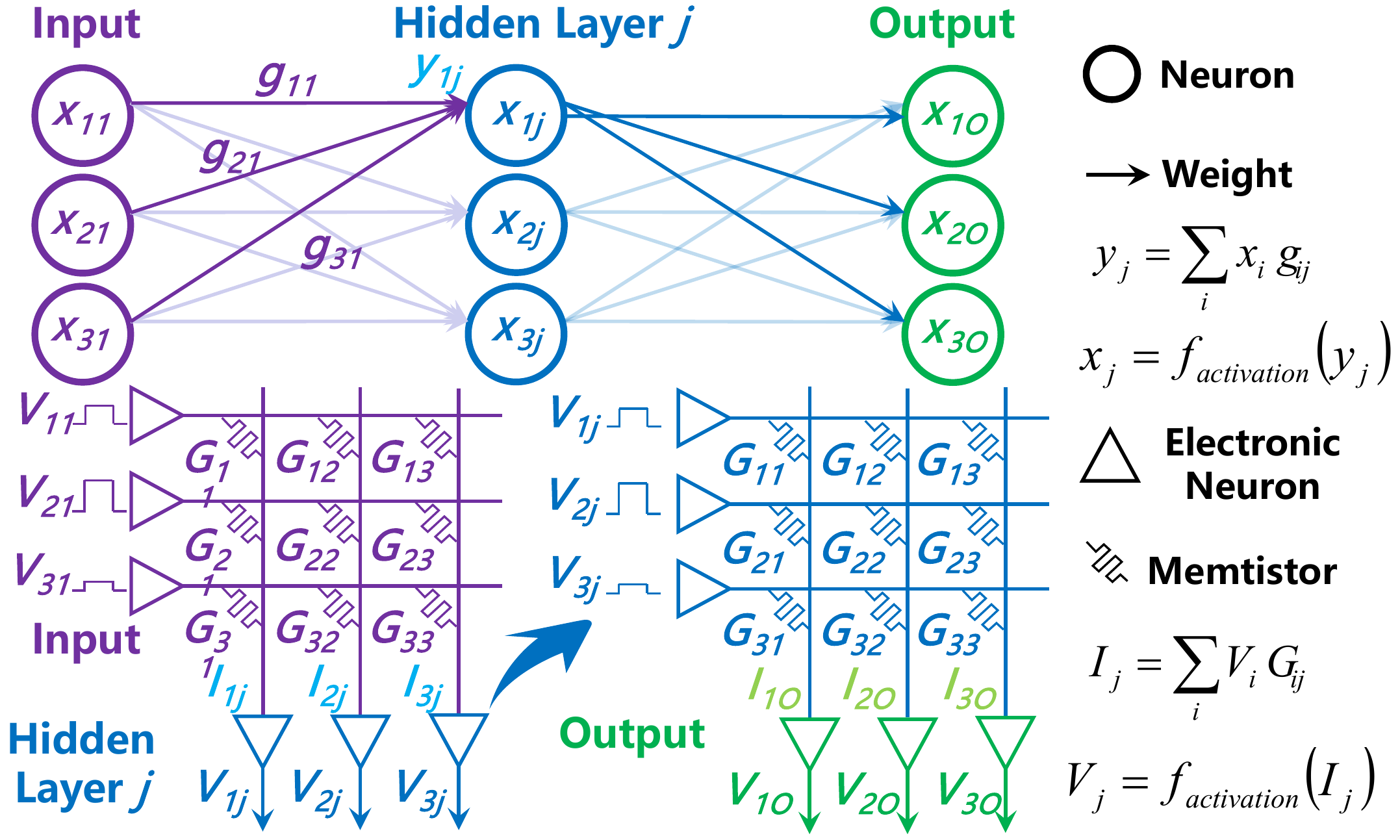}
\caption{Schematic of a memristor based neural network.}
\label{fig1}
\end{figure}

One of the most attractive devices is two-terminal memristor, which offers the advantages of high density, fast speed and low-power consumption \cite{zidan2018future}. The conductance of a memristor can be programmed by update voltage pulses or read by low-amplitude reading voltage pulses \cite{li2018analogue}. Therefore, the weights in memristor based neural networks could be accessed and tuned locally, which is extremely suitable for the DNN hardware representation. As shown in Figure~\ref{fig1}, when implemented in the crossbar structure, memristor arrays are ideal substrates that directly perform multiply-accumulate operations (MACs) at the weight locations. This parallel computing character speeds up the training and inference of DNNs with very low-energy consumption, providing promising computing paradigm for neuromorphic computing systems \cite{yao2017face,zidan2018future,li2018analogue}.

\begin{figure}[htbp]
\centering
\includegraphics[width=0.4\textwidth]{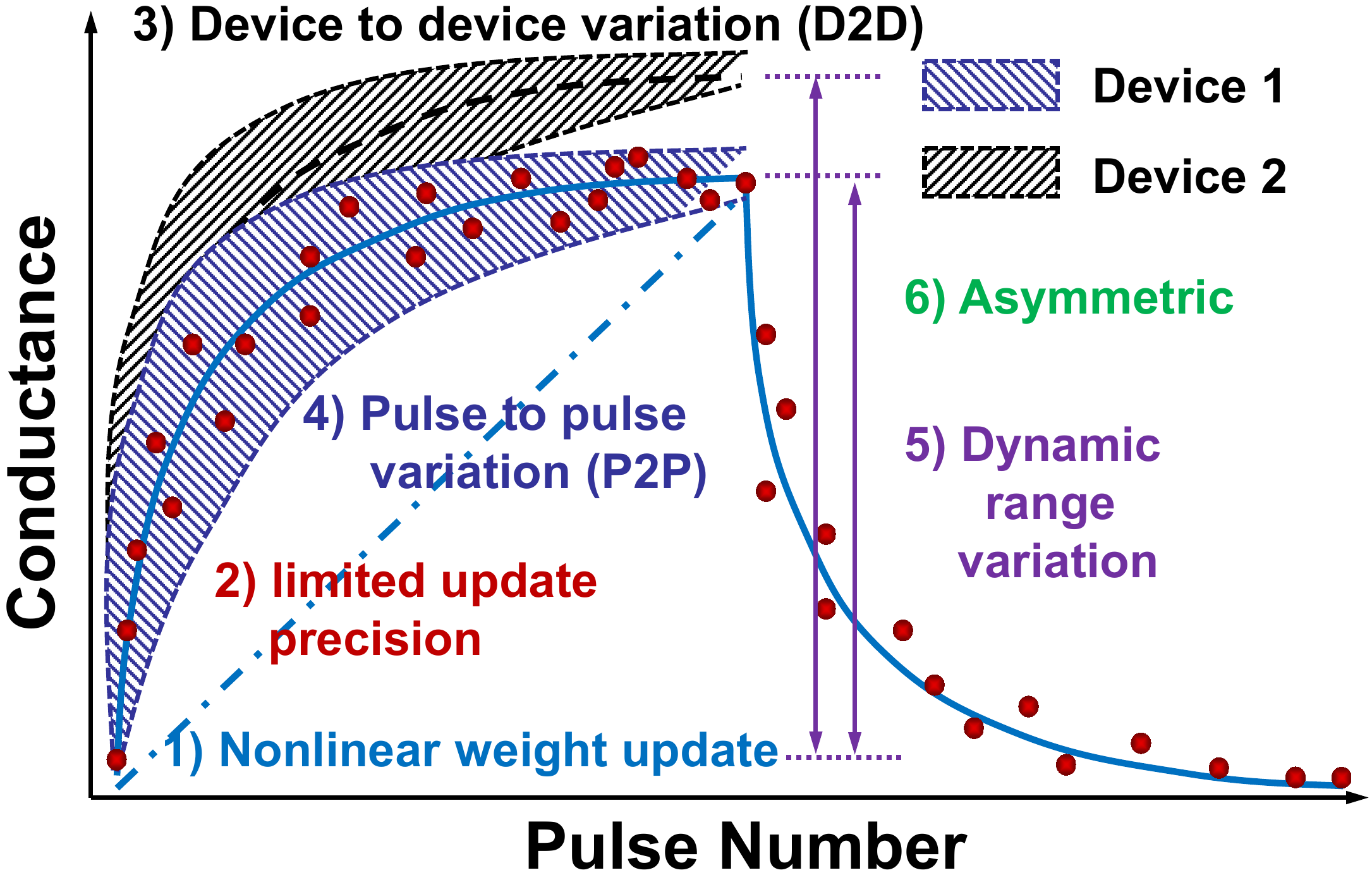}
\caption{Summary of the behaviors in real memristors during weight updating.}
\label{fig2}
\end{figure}

From engineering design perspective, ideal characteristics, including low variability, stable analogue conductance levels and linear updating behaviors, are desired for synaptic memristor devices \cite{yu2018neuro}. However, existing memristor devices have substantial non-ideal characteristics (Figure~\ref{fig2}) \cite{chen2017neurosim+}, such as device-to-device (D2D) variation, pulse-to-pulse (P2P) variation, dynamic range (on/off ratio) variation, and reading noises, which lead to the imprecise encoding of network weights and the corresponding computation. The conductance of memristors is usually determined by the formation and rupture of conductive filaments, which is a relatively stochastic process \cite{yang2013memristive}. Thus, these non-ideal properties are natural and cannot be eliminated at least in the near future. Besides, the updating process of the conductance is nonlinear, asymmetric, and limited-precision (caused by discrete pulse width) \cite{chen2017neurosim+}. These properties discussed above generally lead to non-convergence when training real memristor neural networks.

It is highly desirable to explore an effective solution for the in-situ training on imprecise memristor networks. To this end, we design an efficient and general in-situ training solution for imprecise memristor neuromorphic systems. Our contributions are twofold:
(1)	We build a simulator to quantitatively analyze the effects of various characteristics in memristor based networks. The simulator supports various neural network models and is easy to be inserted into deep learning frameworks. We find that the accuracy degradation is mainly caused by the non-reading updating scheme, pulse width precision, and the D2D variation.
(2)	We propose a learning scheme that integrates stochastic sparse updating with momentum adaption (SSM) to significantly improve the accuracy and learning efficiency of imprecise memristor networks.
The effectiveness of our SSM scheme is demonstrated on various networks and datasets. For MLP on MNIST \cite{lecun1998gradient} dataset, the network accuracy improves from 26.12\% to 90.07\%, and the number of updating pulses decreases 90\%. For CNN on Fashion \cite{xiao2017fashion} dataset, the network accuracy improves from 65.98\% to 92.38\%, and the number of updating pulses decreases 40\%. Meanwhile, the convergence rates of the MLP and the CNN are both 3$\times$ faster.

\section{Related work}
In previous studies, learning schemes that consider all of the non-ideal memristor network properties were rarely reported. The reports of multiple devices for each synaptic weight \cite{yu2015scaling} and additional random sparse adaption (RSA) network \cite{mohanty2017random} were proposed to alleviate the effects of D2D variations. However, these designs increase the circuit overheads and hinder the network size. Some complicated programming schemes were also proposed to improve the precision of weight updates, such as non-identical programming pulses scheme \cite{chen2015mitigating} and closed-loop update scheme \cite{hu2018memristor}. However, the non-identical programming pulses scheme could not be applied with the parallel weight updating method \cite{chen2015technology}, which significantly increases the training latency. The closed-loop update scheme needs several programming cycles to approach the target precision, which extremely increases the number of updating pulses and leads to high latency and energy consumption. Moreover, both schemes need a (or several) read step(s) before weight programming, which inevitably complicate the peripheral circuitry design. A threshold weight update scheme was also proposed to suppress the effects of nonlinear weight changing \cite{chang2018mitigating}, while other memristor characteristics were not considered, such as the variation of D2D and P2P.

\section{Device model}
\begin{figure}[bp]
\centering
\includegraphics[width=0.48\textwidth]{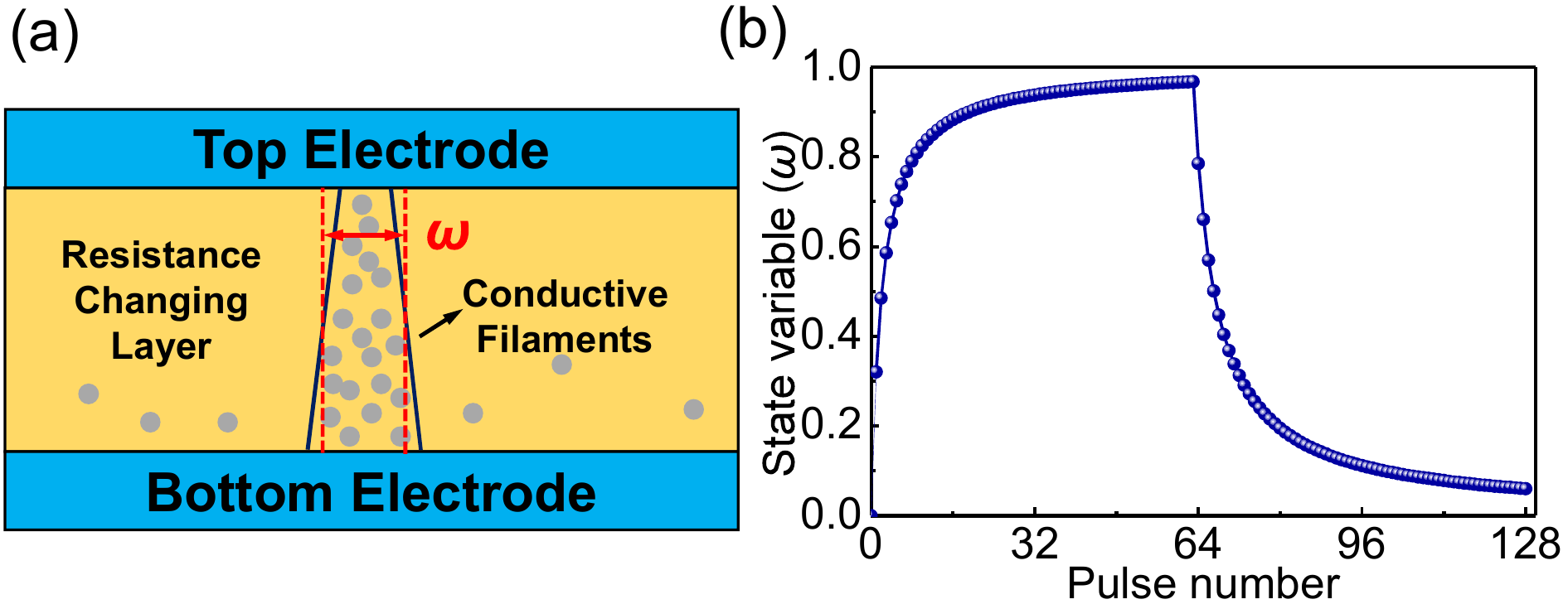}
\caption{(a) Schematic of the memristor. (b) Simulation results of the $\omega$ during 64 potentiation and 64 depression pulses.}
\label{fig3}
\end{figure}

Memristors are usually metal-insulator-metal structured with two-terminals. Their conductance is generally determined by the filaments formed by active metal atoms or oxygen vacancies \cite{yang2013memristive}, as shown in Figure~\ref{fig3}a. The number of the filaments or, equivalently, the area covered by the filaments inside a memristor can be tuned by the external electric field, thus the conductance (weight) of a memristor could be programmed by voltage pulses. In this work, we use a state parameter $\omega \in [0, 1]$ to describe the area covered by filaments in memristors. The dynamic change of $\omega$ in response to the external voltage $V$ is modeled by Equation~\ref{eqn1}, which is verified by the experimental data \cite{choi2015data}:
\begin{equation}
\label{eqn1}
\frac{d\omega}{dt}=
\left\{
    \begin{aligned}
    (1-\omega)^2 k (e^{-\mu_{1}V}-e^{-\mu_{2}V}), V<0, \\
    \omega^2 k (e^{-\mu_{1}V}-e^{-\mu_{2}V}), V>0. \\
    \end{aligned}
\right.
\end{equation}
Where the $k$, $\mu_{1}$, $\mu_{2}$ are positive parameters determined by the material of memristors, $k$ is the ion hopping distance, $\mu_{1}$ and $\mu_{2}$ are the hopping barrier heights.

In Equation~\ref{eqn1}, the exponential dependence dynamic of $V$ and $\omega$ allows memristors to be read at a low amplitude pulse (e.g., bellow 0.1V), which negligibly disturbs their states (the $d \omega /dt$ is small). On the other hand, using high amplitude pulses will program the $\omega$ of the device. During programming process, the potentiation pulses or depression pulses are identical with fixed amplitudes ($V_p$/$V_d$) and widths ($T_p$/$T_d$), which makes the training process simple and suitable for engineering purpose (Table~\ref{tab1}). However, as shown in Figure~\ref{fig3}b, the trade-off is that the device state changes in a nonlinear and discrete (limited pulse width precision) manner.
The current $I$ through the memristor can be modeled \cite{choi2015data} by
\begin{equation}
\label{eqn2}
    I = \omega \gamma {\rm sinh} (\delta  V) + (1-\omega) \alpha (1- e^{-\beta V})
\end{equation}
Where $\gamma$, $\delta$, $\alpha$, $\beta$ are positive parameters determined by the material, $\gamma$ is the effective tunneling distance, $\delta$ is the tunneling barrier, $\alpha$ is the depletion width of the Schottky barrier region and the $\beta$ is the Schottky barrier height. The details of parameters used in the model are shown in Table~\ref{tab1}.

In Equation~\ref{eqn2}, the voltage and current of a memristor is approximately linear-correlated at low voltage (e.g., bellow 0.1V), then the reading conductance is approximately a constant. Thus we chose the reading voltage $V_r$ as 0.05V and the input voltage during learning is below 0.1V.

\begin{figure}[htbp]
\centering
\includegraphics[width=0.48\textwidth]{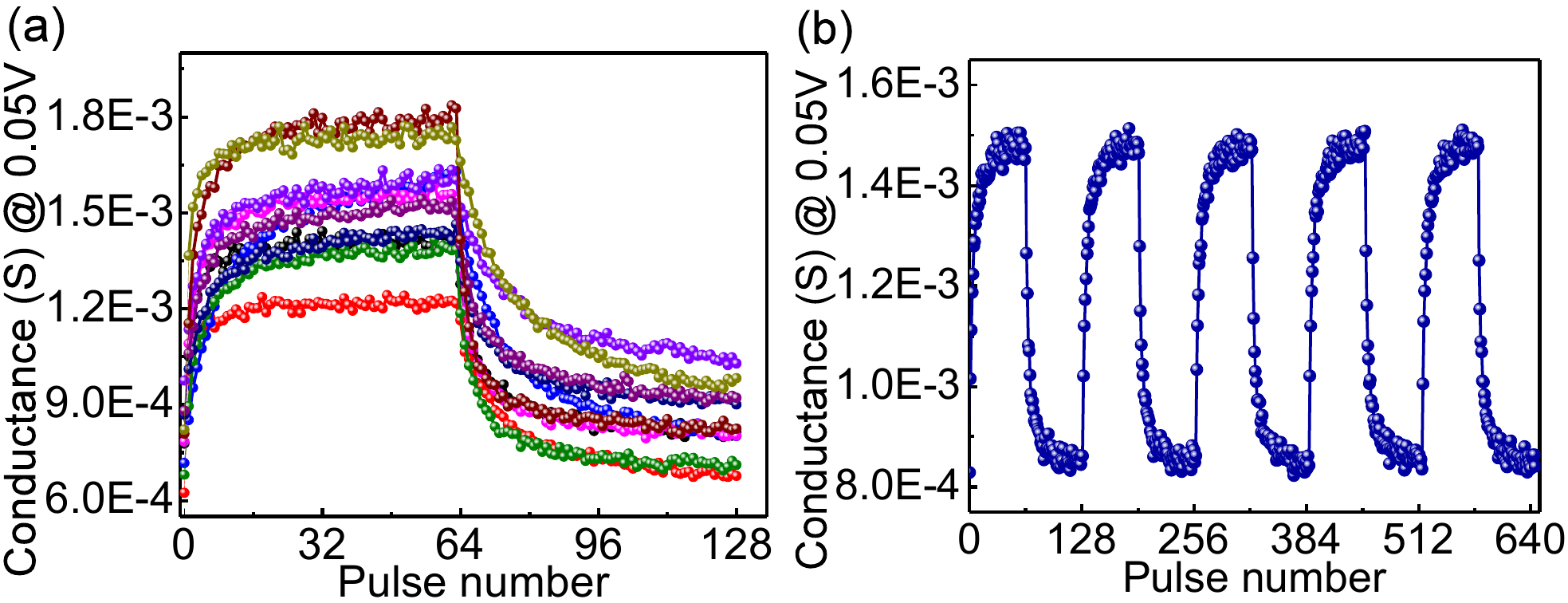}
\caption{Simulation results with all non-ideal properties. (a) D2D variation for 10 devices in one cycle (b) P2P variation for single device in multiple cycles.}
\label{fig4}
\end{figure}

For real memristors, due to fabrication nuances, the conductance update behaviors are different among different devices. To quantify this D2D variation, we assume that the initial parameters in Equations~\ref{eqn1} and \ref{eqn2} of different devices vary according to Gaussian distributions \cite{choi2015data,sheridan2017sparse}. The mean values and standard deviations of each parameter are shown in Table~\ref{tab1}. The standard deviation of each parameter is represented as a percentage of its mean value. Besides the D2D variation, P2P variation is also an important non-ideal property of real memristors. To quantify the P2P variation, we assume that P2P variation is 10\% of the D2D variation. Because of the variations derived from D2D and P2P, the dynamic range (on/off ratio) between maximum and minimum conductance of each device also differs. Figure~\ref{fig4} shows the simulation results with all non-ideal characteristics under voltage pulses.

\renewcommand\arraystretch{1.3}
\begin{table}[htbp]
\caption{List of the parameters used in the memristor model}
\begin{center}
\begin{tabular}{ccc|ccc|cc}
\toprule
Param & mean & std & Param & mean & std & Param & value \\
\hline
$k$ & 1e-4 & 3\% & $\delta$ & 0.5 & 3\% & $V_{\rm p}$ & -1.1V \\
$\mu_1$ & 19.25 & 3\% & $\alpha$ & 1.58e-3 & 15\% & $T_{\rm p}$ & 3 \textmu s \\
$\mu_2$ & 13 & 3\% & $\beta$ & 0.5 & 3\% & $V_{\rm d}$ & 1.4V \\
$\gamma$ & 3.01e-3 & 10\% & $V_{\rm r}$ & 0.05V & - & $T_{\rm d}$ & 30 \textmu s \\

\bottomrule
\end{tabular}
\label{tab1}
\end{center}
\end{table}

\section{Non-ideal effects on memristor networks}

\subsection{Simulator for Non-ideal Memristor Networks}
\begin{figure}[tbp]
\centering
\includegraphics[width=0.48\textwidth]{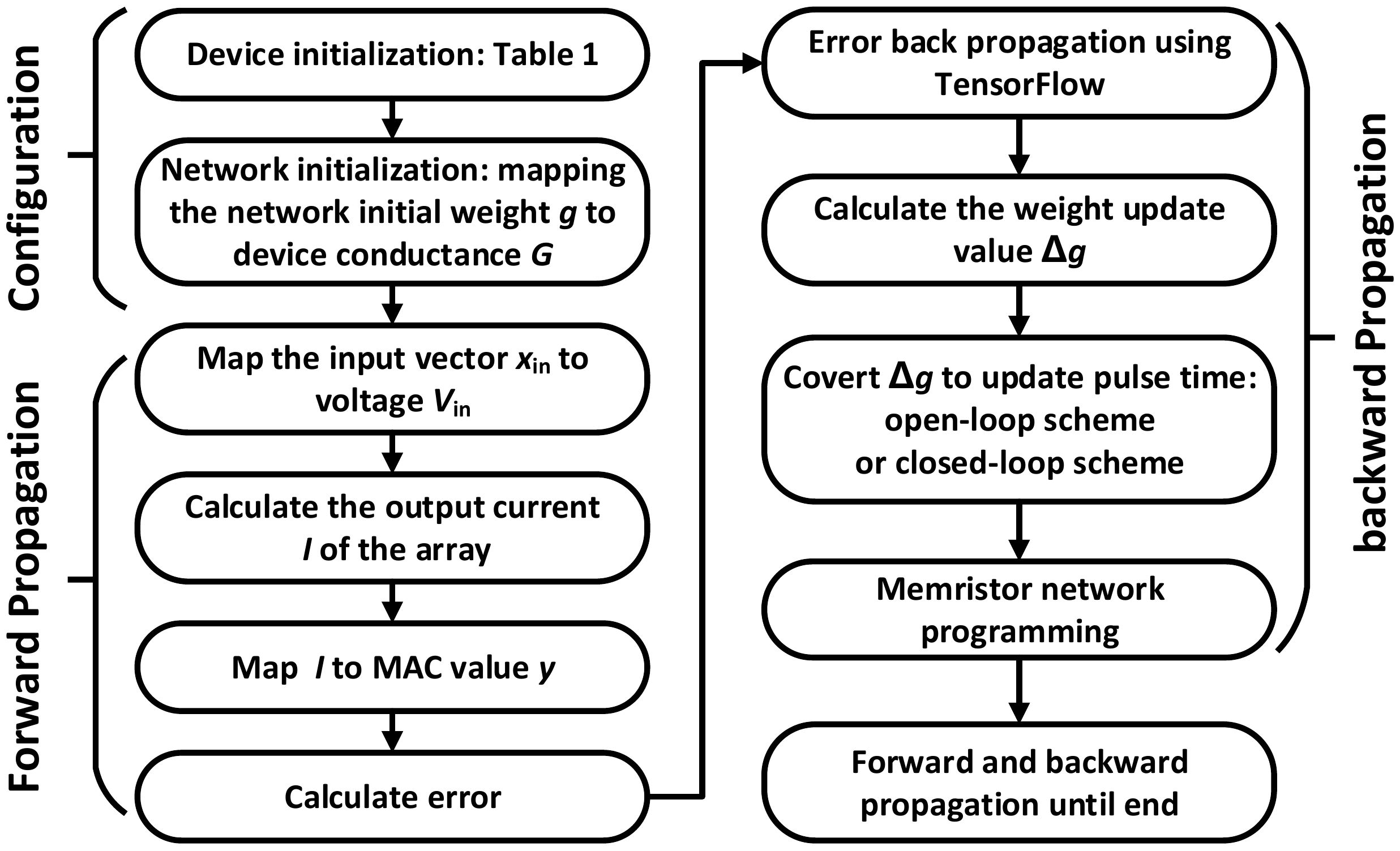}
\caption{Dataflow of inference and training for memristor network.}
\label{fig5}
\end{figure}

As shown in Figure~\ref{fig1}, since MACs can be directly mapped into memristor networks, both the forward and backward process can be performed directly by hardware memristor networks. To quantify the effects of the non-ideal characteristics, we build a one-to-one correspondent simulator based on TensorFlow \cite{abadi2016tensorflow} for representations of memristor networks and evaluations of their performances. The general dataflow during the simulation is shown in Figure~\ref{fig5}.

In our simulation, the parameters of each memristor are defined as collections $P$ and their mean values are defined as $\bar{P}$  (Table~\ref{tab1}). For the network initialization, the $P$ of each memristor among the network are according to a Gaussian distribution. In this condition, each device has its own parameters $P_{ij}$, where $i$ and $j$ denotes the row and column index in memristor array.  Although the initial $\omega$ is same for each device (0.5), individual $P_{ij}$ leads to an approximate Gaussian distribution of the initial conductance among all memristors. The input vector $x_{\rm in}$ is mapped into the amplitude of the voltage pulse $V_{\rm in}$ by
\begin{equation}
\label{eqn3}
    V_{\rm in} = 0.1 x_{\rm in}.
\end{equation}
Since $x_{\rm in}$ is normalized to $[0,1]$, the $V_{\rm in} \in [0,0.1]$, which ensures that $V_{\rm in}$ negligibly disturbs the state of the memristor. The current collected at the output of each column $j$ in the network array can be obtained as:
\begin{equation}
\label{eqn4}
    I_j = \sum_{i} [\omega_{ij} \gamma {\rm sinh} (\delta  V_{i}) + (1-\omega_{ij}) \alpha (1- e^{-\beta V_{i}} ) ] \approx \sum_i V_i G_{ij},
\end{equation}
where the parameters are the $P_{ij}$ of each memristor.

For the hardware representation of MACs, the elements values $g_{ij}$ of a matrix $M$, which represent the weights of the network synapses, are mapped into the conductance values $G_{ij}$ of the memristor:
\begin{equation}
\label{eqn5}
\begin{aligned}
    g_{ij} &= aG_{ij} - b \\
    a &= 2/(\bar{G}_{\rm max} - \bar{G}_{\rm min}), b = (\bar{G}_{\rm max} + \bar{G}_{\rm min}) \\
    \bar{G}_{\rm max} &= [\gamma {\rm sinh} (\delta V_{\rm r})] / V_{\rm r}, \bar{G}_{\rm min} = [\alpha (1-e^{-\beta V_{\rm r}})]  / V_{\rm r} \\
\end{aligned}
\end{equation}
Where $\bar{G}_{\rm max}$ and $\bar{G}_{\rm min}$ are mean values of maximum and minimum conductance. Since the properties of each memristor are not expected to be measured during the mapping, all parameters in Equation~\ref{eqn5} are the mean values of memristor array, i.e., $\bar{P}$ in Table~\ref{tab1}. Then MACs are executed and the output results $y_j$ should be mapped by the current:
\begin{equation}
\label{eqn6}
    y_j = \sum_{i} g_{ij} x_{i} \approx 10 \cdot a I_j - b \sum x_{\rm in}.
\end{equation}

For multi-layer networks, we continue the inference process until the final output layer and calculate the training error. Through the naive error back-propagation algorithm in TensorFlow, the weight update value $\Delta g_{ij}$ corresponding to the mapped $ g_{ij}$ for each memristor is calculated. Once obtaining $\Delta g_{ij}$, the programming pulse time $t_{ij}$ can be calculated. Our simulator has two types weight updating scheme: open-loop and closed-loop. In order to reduce the complexity of peripheral circuits, latency and energy overheads, memristors are expected to be tuned in an open-loop scheme without reading their states:
\begin{equation}
\label{eqn7}
    t_{ij} = n_{ij} \cdot T_{\rm p/d} = \lfloor (N\Delta g_{ij} / 2 ) \rfloor \cdot T_{\rm p/d},
\end{equation}
where $N$ is an amplification coefficient. In previous studies \cite{agarwal2016resistive}, $N$ is generally set as the pulse number required to switch the memristor between minimum and maximum conductance, e.g., 64. For real memristor systems, the programming pulse number $n_{ij}$ should be rounded because of the fixed programming pulse width.

On the other hand, the closed-loop scheme provides a relatively precise updating. In our simulation configuration, the closed-loop updating requires just one reading step before programming to check the memristor conductance $G_{ij}$:
\begin{equation}
\label{eqn8}
  G_{ij} = [\omega_{ij} \gamma {\rm sinh} (\delta  V_{\rm r}) + (1-\omega_{ij}) \alpha (1- e^{- \beta V_{\rm r}}) ] / V_{\rm r}.
\end{equation}
The state $\omega_{ij}$ of the memristor is
\begin{equation}
\label{eqn9}
    \omega_{ij} = \frac{ G_{ij} V_{\rm r} - \alpha (1- e^{-\beta V_{\rm r}})}   {a [\gamma {\rm sinh} (\delta V_{\rm r}) - \alpha (1-e^{-\beta V_{\rm r}})]}.
\end{equation}
The expected update state $\Delta \omega_{ij}$ of the memristor is:
\begin{equation}
\label{eqn10}
    \Delta \omega_{ij} = \frac{\Delta g_{ij} V_{\rm r}}  {a [\gamma {\rm sinh} (\delta V_{\rm r}) - \alpha (1-e^{-\beta V_{\rm r}})]}.
\end{equation}
Therefore, the programming pulse time $t_{ij}$ can be obtained as
\begin{equation}
\label{eqn11}
\begin{aligned}
t &= \lfloor \frac{\Delta \omega_{ij}}{\lambda k (e^{-\mu_{1} V_{\rm p/d}} - e^{\mu_{2} V_{\rm p/d}})(\lambda - \Delta \omega_{ij}) T_{\rm p/d}} \rfloor \cdot T_{\rm p/d} \\
\lambda &= \left\{
    \begin{aligned}
    (1-\omega_t), V=V_p. \\
    -\omega_t, V=V_d. \\
    \end{aligned}
\right.
\end{aligned}
\end{equation}

It should be pointed out that, due to the reading step, the parameters in Equation~\ref{eqn8} is the $P_{ij}$ of each memristor. As for the next Equations~\ref{eqn9}-\ref{eqn11}, the parameters are the mean value of each parameter $\bar{P}$ in Table~\ref{tab1} due to the fact that the properties of each memristor are not measured during learning.

Finally, the memristors are programmed by the voltage pulses. The dynamics of the memristor are determined by Equation~\ref{eqn1}. To introduce the P2P variation into the model, the parameters $P_{ij}$ of one memristor randomly change under a Gaussian distribution before each pulse.

\subsection{Ablation Studies of Non-Ideal Characteristics}
We test the aforementioned simulator with various optional characteristics in real memristor-based MLP and CNN. The MLP has a 3-layer structure with 784 input neurons, 256 hidden neurons and 10 output neurons. The CNN has a LeNet-5 structure \cite{lecun1998gradient} with two convolution layers and two fully-connected layers. We use the MNIST and Fashion datasets to test the network performance of MLP and CNN, respectively.

\begin{figure}[htbp]
\centering
\includegraphics[width=0.48\textwidth]{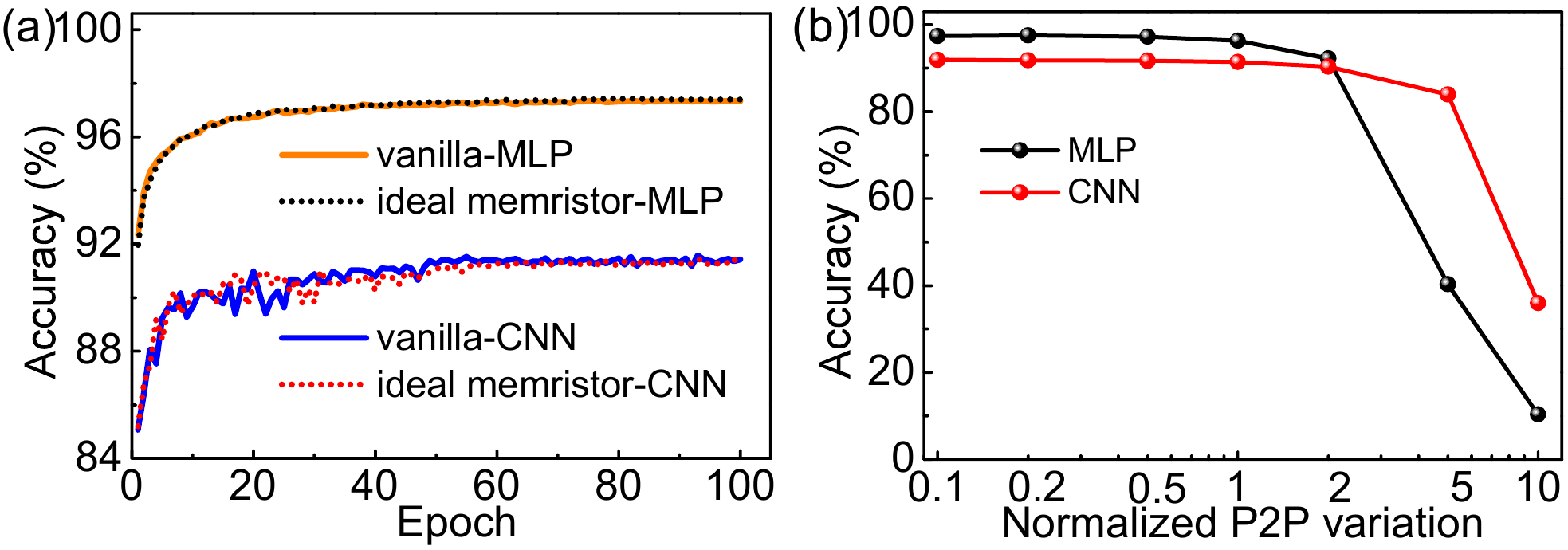}
\caption{(a) Network accuracy with unlimited and limited weight ranges by vanilla networks and ideal memristor networks, respectively. (b) Effects of the P2P variation on network accuracy.}
\label{fig7}
\end{figure}

Because of the on/off ratio of memristors, the weight values mapped by them are also limited. In our model, the expectations of maximum and minimum conductance are mapped into the weight +1 and -1 (Equation~\ref{eqn5}), respectively. While for vanilla networks based on software platform, the synaptic weight have no range limitation. We evaluate the effects of limited weight range on the non-variation memristor network. In this condition, the variation of D2D and P2P are ignored so that the mapped weights only range in [-1, +1], and the weight update scheme is closed-loop. As shown in Figure~\ref{fig7}a, ablation experiments indicate that the range limitation has little effect on the network performance for both MLP and CNN.

For the P2P variation, we chose the parameters in Table~\ref{tab1} as a baseline (existing devices), and zoomed them exponentially to study the effects on the network. In this condition, the weight update scheme is closed-loop and the variation of D2D is ignored. As shown in Figure~\ref{fig7}b, the MLP can tolerate up to 2$\times$ P2P variation baseline and the CNN can tolerate up to 5$\times$. This is contributed to the high robustness of neural networks. We conclude that the single P2P variation is not a major factor for the memristor network accuracy degradation.

\begin{figure}[htbp]
\centering
\includegraphics[width=0.48\textwidth]{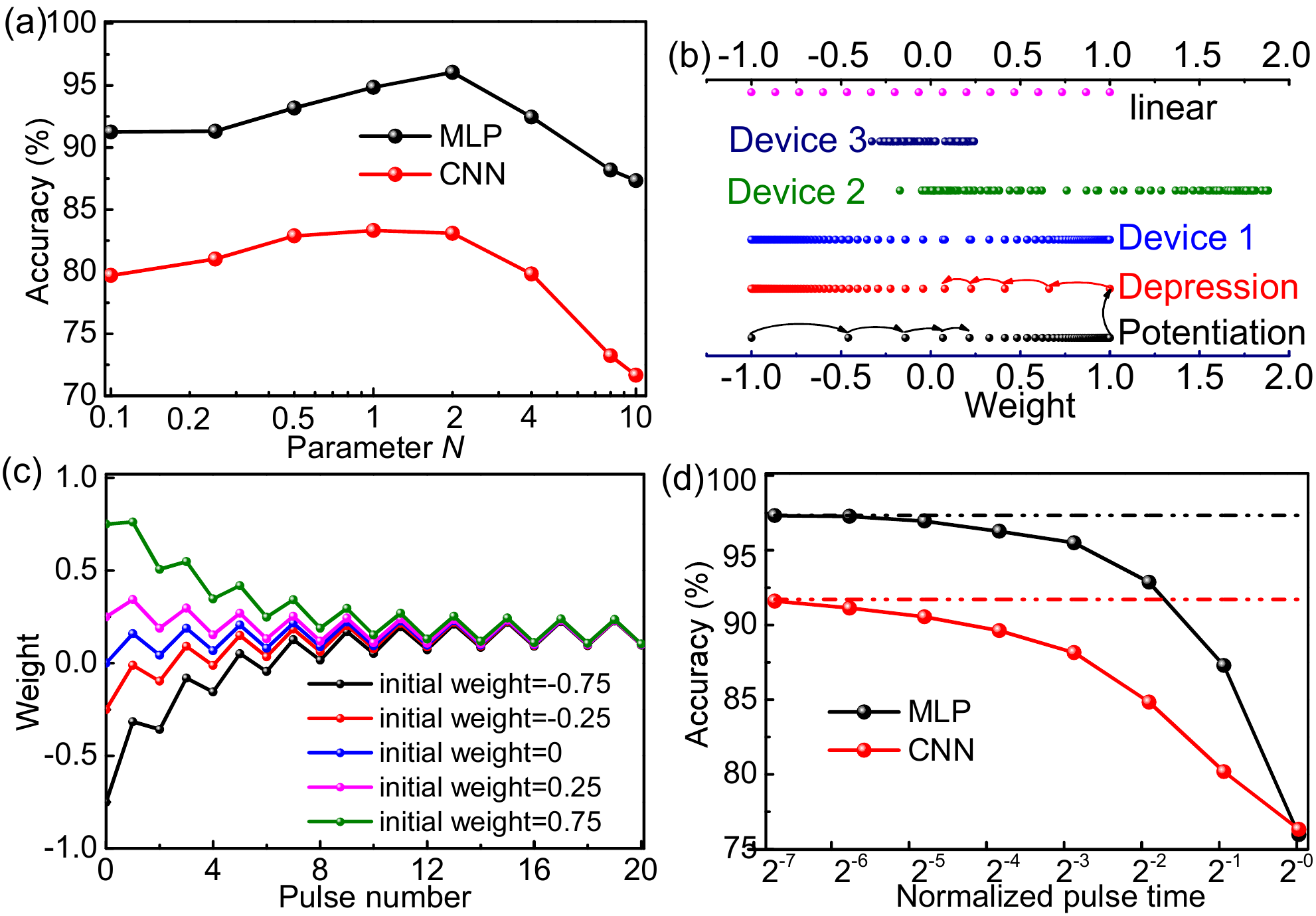}
\caption{(a) Effects of the update coefficient $N$ on the network accuracy. (b) Schematic of the memristor weight distribution with discrete pulses and D2D variation. (c) Evolution of the weight under identical alternating potentiation and depression pulses. (d) Effects of the pulse width on the network accuracy, the dash-dot line denotes the accuracy in ideal memristor network.}
\label{fig8}
\end{figure}

For practical online learning applications of memristor networks, an open-loop weight update scheme is much more attractive than the closed-loop scheme due to its non-reading and fast update properties. We studied the effect of the update coefficient $N$ on the network performance in the open-loop scheme. In this condition, the update pulses are rounded to integral numbers, and the variation of D2D and P2P are ignored. As shown in Figure~\ref{fig8}a, the results show that a relatively small $N$ (e.g., 2) can improve the accuracy in both MLP and CNN, rather than setting it as the pulse number required to switch the memristor between the minimum and maximum conductance (e.g., 64). Through Equation~\ref{eqn7}, we find that coefficient $N$ plays a similar role as the learning rate during training. Like the learning rate, an overlarge $N$ may lead to an unstable training process or even non-convergence. On the other hand, a local minimum result occurs if $N$ is too small, which is also undesirable for the network performance. Meanwhile, the network accuracy still decays by a large margin even after the optimization. This is due to the inaccurate updates during the open-loop scheme, especially when applying the pulse number rounding. As shown in Figure~\ref{fig8}b, due to the discrete pulse number, the impact of a depression (potentiation) pulse is widely different from a potentiation (depression) one at the maximum (minimum) conductance. This asymmetric behavior is further detailed in Figure~\ref{fig8}c, where the weight decays towards a center value under identical alternating potentiation and depression pulses. This nonlinear and asymmetric weight update properties seriously degrade the accuracy of the network. A shorter programming pulse width ($T_{\rm p/d}$) can reduce this effect. We chose the pulse widths in Table~\ref{tab1} as a baseline (existing devices, $2^0$ in Figure~\ref{fig8}d), and narrowed them exponentially to study the effects on the network. As shown in Figure~\ref{fig8}d, a shorter pulse width results in a higher accuracy. However, if the programming pulse is too short, the state of a memristor would not change, which is called the delay time \cite{zidan2018future}. Thus, the pulse width is not supposed to be less than that in baseline.

\begin{figure}[htbp]
\centering
\includegraphics[width=0.3\textwidth]{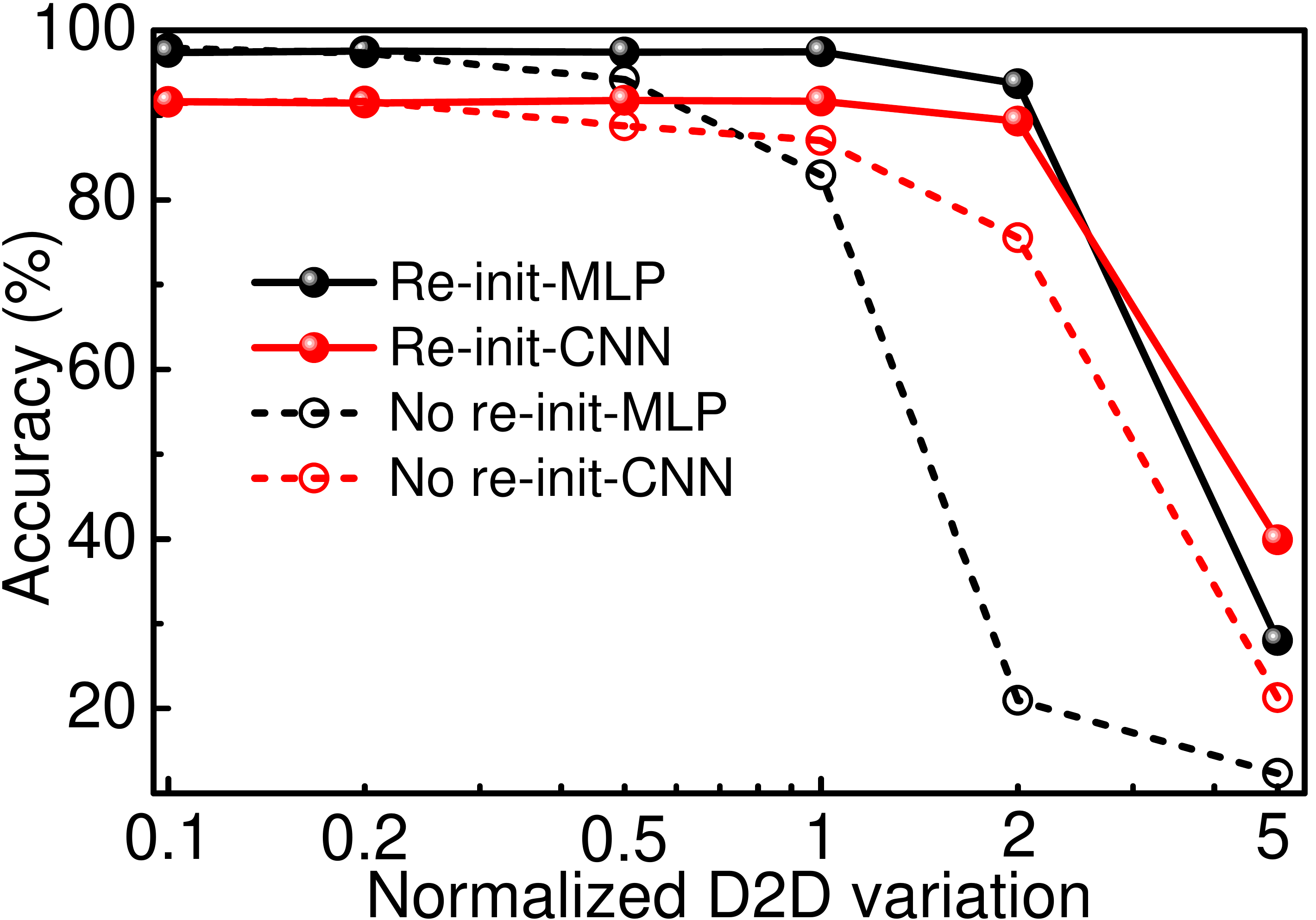}
\caption{Effects of D2D variation and re-initializing on the network accuracy.}
\label{fig9}
\end{figure}

For the D2D variation, we chose 10\% of the D2D variation in Table~\ref{tab1} as a baseline (existing devices), and exponentially zoomed them to study the effects on the network. In this condition, the weight update scheme is closed-loop and the variation of P2P is ignored. As shown in Figure~\ref{fig9}, both the MLP and the CNN exhibit tolerances to the D2D variation, which are weaker than the P2P variation in Figure~\ref{fig7}b. We find that the accuracy degradation is mainly caused by the weight initialization, which is crucial for training DNNs \cite{mishkin2015all}. D2D variation broadens initial weight distribution, thus resulting in an improper starting point of the network training. To eliminate this effect, we design a re-initialization method (see Section~\ref{re_init}). With the re-initialization, both the MLP and the CNN can be further enhanced to tolerate up to 2$\times$ D2D variation baseline. However, if the D2D is too large (e.g., 5$\times$ D2D variation baseline), the network accuracy would also be significantly declining. Since the D2D variation of existing devices is close to the baseline, our re- initialization method is powerful enough.

Considering all the non-ideal characteristics discussed above, the network accuracy decreases to extremely low levels or even non-convergence: 26.12\% for MLP and 65.98\% for CNN. As summarized in Table~\ref{tab2}, through the comparisons of accuracy drops in the above ablation studies, we conclude that the degradation in real imprecise memristor networks is caused by three major factors: the open-loop update scheme, the pulse width precision, and the D2D variation.

\renewcommand\arraystretch{1.3}
\newcommand{\tabincell}[2]{\begin{tabular}{@{}#1@{}}#2\end{tabular}}
\begin{table}[htbp]
\caption{The simulation results of the MLP and CNN with different non-ideal characteristics.}
\begin{center}
\begin{tabular}{c|c|c|c|c|c|c}
\toprule
\multicolumn{2}{c|}{Variation} & \multirow{2}{*}{\tabincell{c}{Open \\ -loop}} & \multirow{2}{*}{\tabincell{c}{Roun \\ ding}} &
\multicolumn{2}{c|}{Test accuracy (\%)} & \multirow{2}{*}{Comment} \\
\cline{1-2}
\cline{5-6}
P2P & D2D & & & MLP & CNN & \\
\hline
$\times$ & $\times$ & $\times$ & $\times$  & 97.44 & 91.66 & vanilla \\
\hline
$\times$ & $\times$ & $\times$ & $\times$  & 97.34 & 91.71 & ideal \\
\hline
\checkmark & $\times$ & $\times$ & $\times$ & 97.43 & 91.64 & tolerant \\
\hline
$\times$ & \checkmark & $\times$ & $\times$ & \textbf{82.93} & \textbf{86.99} & \multirow{3}{*}{\tabincell{c}{ \textbf{major}\\ \textbf{factors}}}\\
\cline{1-6}
$\times$ & $\times$ & \checkmark & $\times$ & 96.05 & \textbf{83.09} & \\
\cline{1-6}
$\times$ & $\times$ & $\times$ & \checkmark & \textbf{75.96} & \textbf{76.32} & \\
\hline
\checkmark & \checkmark & \checkmark & \checkmark & \textbf{26.12} & \textbf{65.98} & non-ideal \\

\bottomrule
\end{tabular}
\label{tab2}
\end{center}
\end{table}

\section{Stochastic sparse learning with momentum adaptation (SSM)}
Based on the simulation results, we design an SSM scheme to address these issues in the non-ideal memristor network training and enable the network to converge with high performance. The SSM scheme are decomposed as follows.

\subsection{Re-initialization}
\label{re_init}
A suitable uniform or Gaussian weight initialization with proper variation is benefit for the neural network forward and backward propagation. Thus, we design a simple re-initialization method to reduce the effect of the D2D variation on the network initialization, as shown in Table~\ref{tab3}.

\renewcommand\arraystretch{1.3}
\begin{table}[htbp]
\caption{Re-initialization of the memristor network.}
\begin{center}
\begin{tabular}{c|l|c}
\toprule

\textbf{1} &  \multicolumn{2}{l}{Read memristor conductance $G$, map it to network weight $g$.} \\
\hline

\multirow{6}{*}{\textbf{2}} & \multirow{6}{*}{\tabincell{c}{ Gaussian re-initialization:\\ \\ \textbf{if} $g_{ij} >= 0$ \\ apply a depression pulse \\ \textbf{else} \\ apply a potentiation pulse }} & \multirow{6}{*}{\tabincell{c}{ Uniform re-initialization:\\ $\epsilon$ is the boundary (e.g., 0.1) \\ \textbf{if} $g_{ij} >= \epsilon$ \\  apply a depression pulse \\ \textbf{else if}  $g_{ij} <= -\epsilon$\\ apply a potentiation pulse }} \\
& &  \\
& &  \\
& &  \\
& &  \\
& &  \\

\hline
\textbf{3} & \multicolumn{2}{l}{Read and calculate the standard deviation of all synaptic weights.} \\
\hline
\textbf{4} & \multicolumn{2}{l}{Repeat \textbf{2} and \textbf{3} until suitable standard deviation occurs.} \\

\bottomrule

\end{tabular}
\label{tab3}
\end{center}
\end{table}

\subsection{Momentum Based Gradient Method}
Due to the non-ideal characteristics of memristor networks, training based on the naive stochastic gradient descent (SGD) usually leads to non-convergence.
Therefore, we introduce a momentum $m$ into the SGD, named as momentum-based SGD (m-SGD) \cite{krizhevsky2012imagenet} in the SSM scheme. As shown in Figure~\ref{fig10}a, each synaptic cell keeps an exponentially weighed average of all previous gradients during training, such that:
\begin{equation}
\label{eqn13}
\begin{aligned}
    m_{t+1} &= vm_{t} + (1-v)\Delta g_{t}, \\
    g_{t+1} &= g_{t} - \eta m_{t+1}.
\end{aligned}
\end{equation}
Where $\eta$ is the learning rate, $v$ is the momentum coefficient defined in $[0,1]$. This averaged gradient update can smooth the weight evolution process by filtering out harmful noises that contains little information during one iteration, and distills robust updates that indicate the right directions, which leads to faster and better convergence during training.

\subsection{Stochastic Rounding Method}
When the $N$ is small, e.g., 2, as suggested by the experiment results in Figure~\ref{fig8}a, the programming pulse number $n$ for each cell in one iteration is scaled to a small order of magnitude, e.g., $[0.01, 0.1]$ and later smoothed by momentum $m$. Then the pulse width is usually too narrow for practical implementation. In the SSM scheme, as shown in Figure~\ref{fig10}b, we apply a stochastic rounding method with ${\rm Bernoulli}$ \{0, 1\} to quantize the pulse number \cite{wu2018training}, otherwise a deterministic rounding will always result in zero pulse:
\begin{equation}
\label{eqn14}
\left\{
    \begin{aligned}
    p_{\rm one~pulse} &= n \\
    p_{\rm no~pulse} &= 1-n\\
    \end{aligned}
\right.
\end{equation}
Where $p$ denotes the probability of applying one update pulse on the memristor. The small order of magnitude $N$ makes the update number $n$ extremely sparse, where only relatively significant weight update $\Delta g$ can affect a pulse during training.

\subsection{Compensation Update Method}
\begin{figure}[htbp]
\centering
\includegraphics[width=0.4\textwidth]{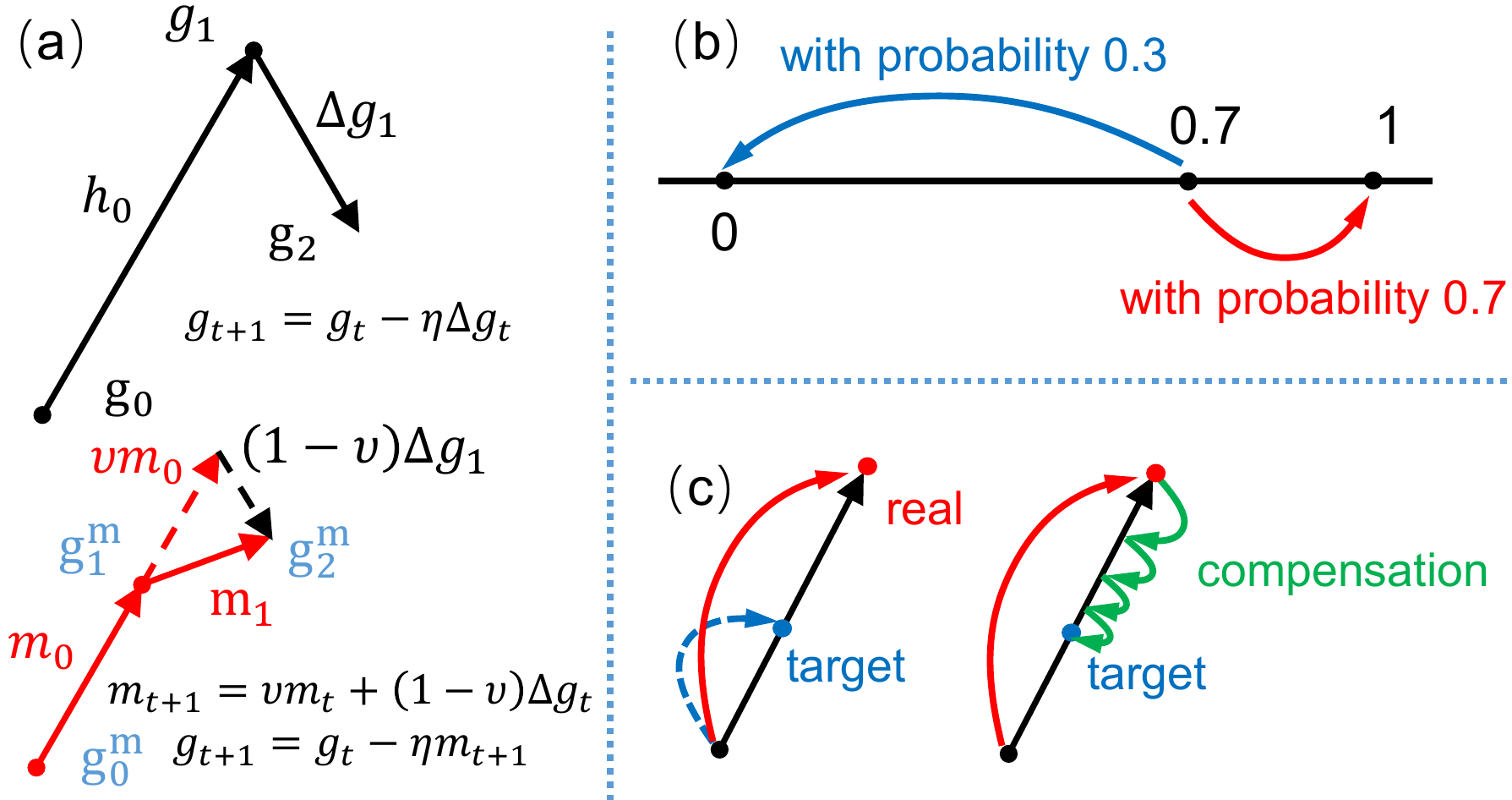}
\caption{Schematic of (a) momentum based gradient descent, (b) stochastic rounding, and (c) compensation update.}
\label{fig10}
\end{figure}

To further mitigate the asymmetric update behavior and improve the precision of the weight update, a compensation method is designed when weights are near the maximum or minimum conductance (Figure~\ref{fig10}c). The current state $\omega_{ij}$, the expected update state $\Delta \omega_{ij}$ of the memristor and the $\lambda$ can be calculated by Equations~\ref{eqn9}-\ref{eqn11}. Then the compensation programming pulse time $t_c$ can be obtained as:
\begin{equation}
\begin{aligned}
    t_c &= \lfloor \frac{ \Delta\omega_{ij} + \psi (\Delta \omega_{ij}-\lambda)  }{ \xi\lambda (\psi + 1)(\lambda-\Delta \omega_{ij})  }   \rfloor \cdot T_{p/d} \\
    \xi &= k(e^{-\mu_{1}V_{p/d}} - e^{\mu_{2}V_{p/d}}), \psi = T_{p/d}\xi\lambda
\end{aligned}
\end{equation}
Obviously, the trade-off is that the compensation needs extra pulses and at least one read operation to calculate its number. Therefore, the compensation method is optional in SSM when the application requires higher accuracy.

\section{Result and discussion}

\subsection{Performance of the Re-initialization Method}
\begin{figure}[htbp]
\centering
\includegraphics[width=0.48\textwidth]{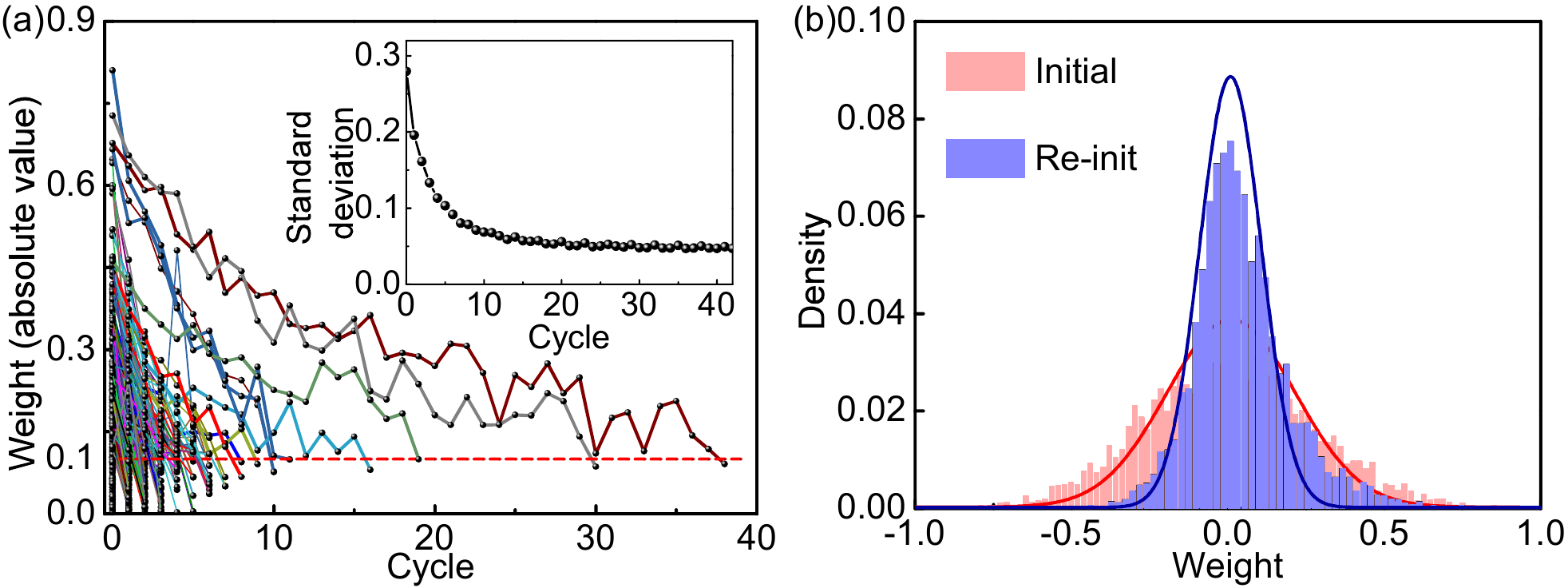}
\caption{(a) Evolution of weights and their standard deviation (insert panel) under uniform re-initialization. (b) Distributions of the weight before and after Gaussian re-initialization.}
\label{fig11}
\end{figure}

As shown in Figure~\ref{fig11}, by our re-initialization, the conductance of the memristor network can be tuned to a suitable narrow deviation within 40 cycles for both uniform and Gaussian scheme. And the average pulse number applied on each cell is about 4. These results indicate that the re-initialization is fast and energy-efficient, and the effects of the D2D variation can be mitigated (Figure~\ref{fig9}).

\subsection{Performance of SSM the Scheme for MLP and CNN}
\begin{figure}[htbp]
\centering
\includegraphics[width=0.3\textwidth]{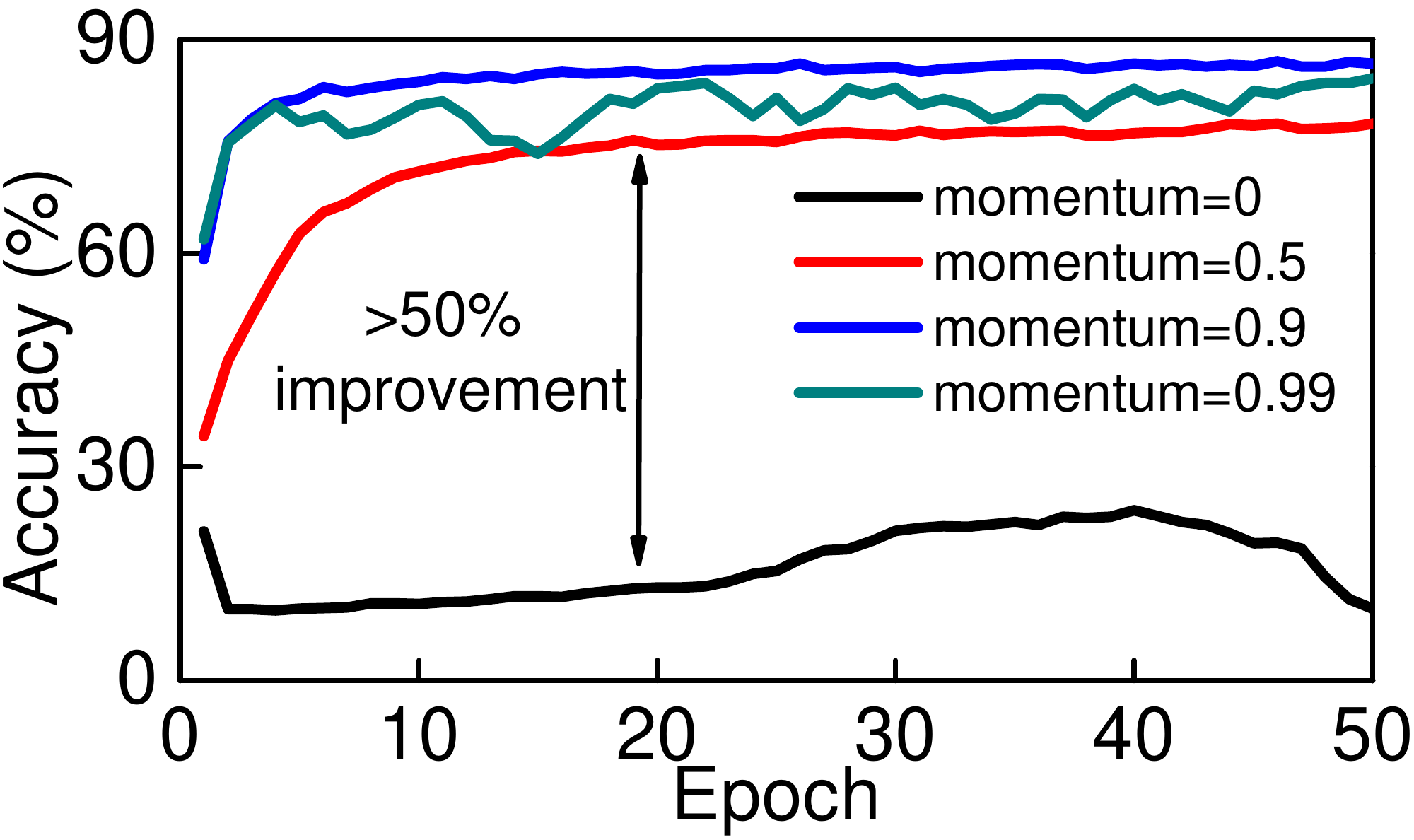}
\caption{Training curves of Memristor based MLP with SSM scheme and without compensation.}
\label{fig12}
\end{figure}

SSM significantly improves the network online learning, as shown in Figure~\ref{fig12}. With all of the memristor non-ideal characteristics, the network cannot converge (black line), while the network can quickly converge to a high accuracy (87.18\%) with SSM scheme and a suitable momentum coefficient $0.9$ (blue line). We also find that an overlarge $v$ (e.g. 0.99) might make the training process unstable and decrease the performance, which has similar phenomenon as the update coefficient $N$. The accuracy can further improve to 90.07\% if compensation is applied (Table~\ref{tab4}).

\begin{figure}[htbp]
\centering
\includegraphics[width=0.48\textwidth]{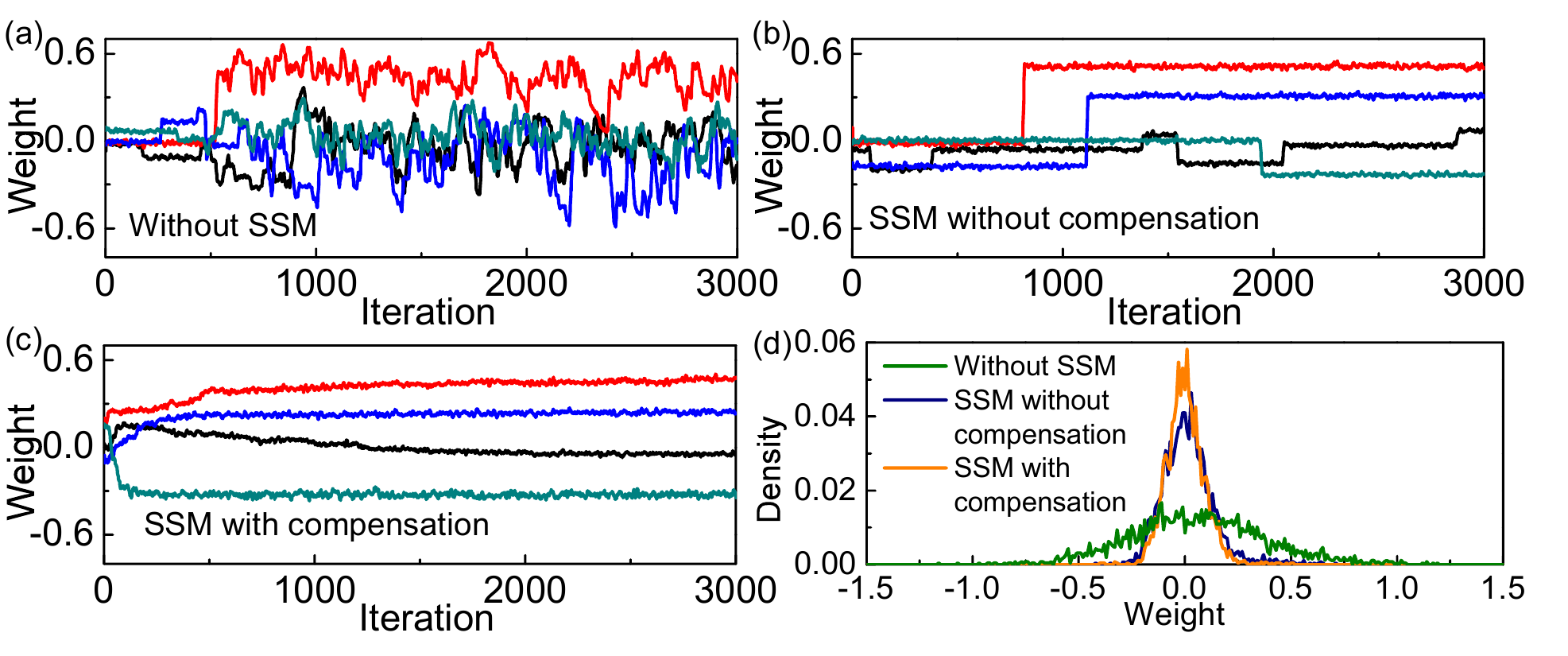}
\caption{(a-c) Three typical synaptic weight update evolutions during online learning: (a) without SSM, (b) SSM without compensation and (c) SSM with compensation. (d) The weight distribution after learning under different schemes.}
\label{fig13}
\end{figure}

We also extract typical synaptic weight update evolutions during training to visualize the role of the SSM scheme. As shown in Figure~\ref{fig13}a, without SSM scheme, even though small updates (noises) can be cleared by the deterministic pulse rounding, the evolution of the weight still fluctuates seriously and leads to an extreme unstable training of the network. We believe this phenomenon is caused by the temporary and local learning information that is constantly changing in each iteration, which contains little explicit direction. As shown in Figure~\ref{fig13}b, we find that the SSM scheme with a suitable momentum $v$ can obviously decrease the programming pulses for each cell during training. Statistical results of the first 5 epochs, which contains most update events, show that less than 1\% of cells are updated in one iteration. In the following epochs, very few devices need to be tuned. This sparse update can highly decrease the energy consumption, as well as stabilize the training and improve the final accuracy (Figure~\ref{fig12}). This is due to the fact that SSM scheme can distill robust updates from temporary information in one iteration and indicate the right directions of updating, which leads to a faster and better convergence.

Due to the limited pulse width precision, the synaptic weight still changes sharply at the update point (Figure~\ref{fig13}b), which usually indicates an imprecise updating and may hurt the overall network performance. As shown in Figure~\ref{fig13}c, with the compensation update method, the weight changes more precisely and smoothly during training, which can further improve the network performance at the cost of reading operations and more pulses.

As shown in Figure~\ref{fig13}d, we see that without the SSM scheme, the distribution of the weight after learning is dispersed, while the SSM scheme can narrow the weight distribution and avoid the non-convergence. All of these behaviors discussed above are also verified in the CNN training.

We summarize performance details of MLP and CNN network with the SSM scheme in Table~\ref{tab4}. With the SSM scheme, the classification accuracy increases from 26.12\% to 87.18\% in MLP, and the programming pulse number in the first 5 epochs decreases 90\%. As for the CNN, the classification accuracy increases from 65.98\% to 90.99\%, and the programming pulse number decreases 40\%. The convergence rates of MLP and CNN are both 3 times faster. When the compensation method is applied, the classification accuracy can further reach to 90.07\% and 92.38\% for MLP and CNN, respectively, while the pulse number still approximates to that without the SSM scheme. Therefore, we conclude that our SSM scheme can significantly improve the performance of non-ideal memristor network.

\renewcommand\arraystretch{1.3}
\begin{table}[htbp]
\caption{Summary of SSM experiments for MLP and CNN.}
\begin{center}
\begin{tabular}{c|c|c|c|c}
\toprule

& \multirow{2}{*}{Scheme} & \multirow{2}{*}{\tabincell{c}{ Accuracy \\  (\%)}}
& \multirow{2}{*}{\tabincell{c}{ Pulse \\ count*}} & \multirow{2}{*}{Epoch} \\
&        &               &             &       \\
\hline

\multirow{3}{*}{\tabincell{c}{ M\\ L \\ P}}
& No SSM & 26.12 & 1 & 25 \\
\cline{2-5}
& SSM without compensation & \textbf{87.18} & 0.1 & \multirow{2}{*}{\textbf{8}} \\
\cline{2-4}
& SSM with compensation & \textbf{90.07} & 0.8 & \\
\hline
\multirow{3}{*}{\tabincell{c}{ C\\ N \\ N}}
& No SSM & 65.98 & 1 & 50 \\
\cline{2-5}
& SSM without compensation & \textbf{90.99} & \textbf{0.6} & \multirow{2}{*}{\textbf{16}} \\
\cline{2-4}
& SSM with compensation & \textbf{92.38} & 1.1 & \\

\bottomrule

\multicolumn{5}{l}{*The pulse counts are normalized by the value of no-SSM scheme.}

\end{tabular}
\label{tab4}
\end{center}
\end{table}

\subsection{Performance of the Compensation Update Method}
We evaluated the performance of the compensation update method for different $\omega$ values by stochastically choosing 1000 depression and potentiation weights. The range of $\omega$ is $[0, 1]$. And the range of the update weight is $[-0.1, 0.1]$, which covers most conditions during the network learning.

\begin{figure}[bp]
\centering
\includegraphics[width=0.48\textwidth]{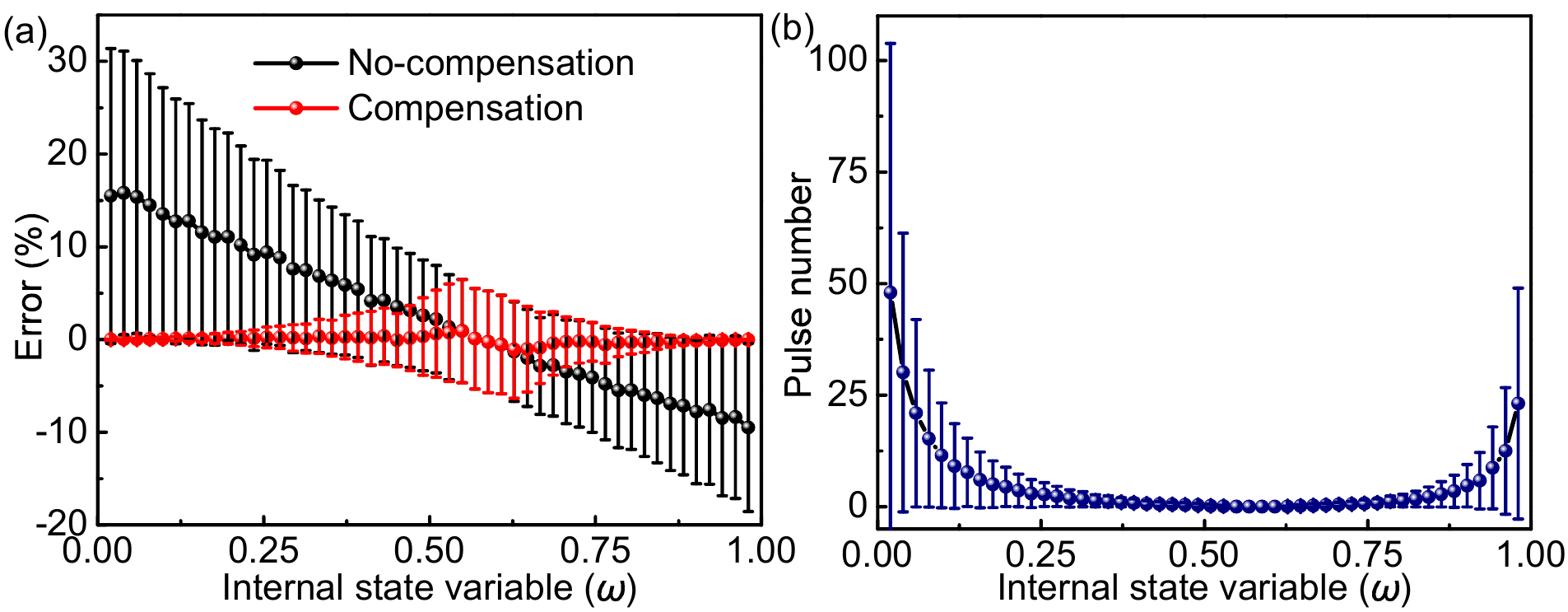}
\caption{Average (a) weight update errors (dots) and (b) compensation pulse number (dots) with compensation method. Bars represent standard deviations.}
\label{fig14}
\end{figure}

As shown in Figure~\ref{fig14}, with the compensation, the average error of the weight update is reduced and the update precision is improved. At the maximum or minimum conductance condition ($\omega=0$ or $1$), the number of compensation pulse is much higher due to the asymmetric update behaviors of the device: when the weight is near maximum (or minimum), a depression (or potentiation) update pulse significantly decreases (or increases) the weight, while a potentiation (or depression) compensation pulse does not change the weight much. Although this large number compensation improves the precise of the weight updating, the trade-off is that the energy cost and the delay time are significantly increased, thus the compensation method is suitable for the application that is insensitive for the efficiency but demands for the high accuracy.

\subsection{Prospects of the SSM Scheme}
\begin{figure}[htbp]
\centering
\includegraphics[width=0.48\textwidth]{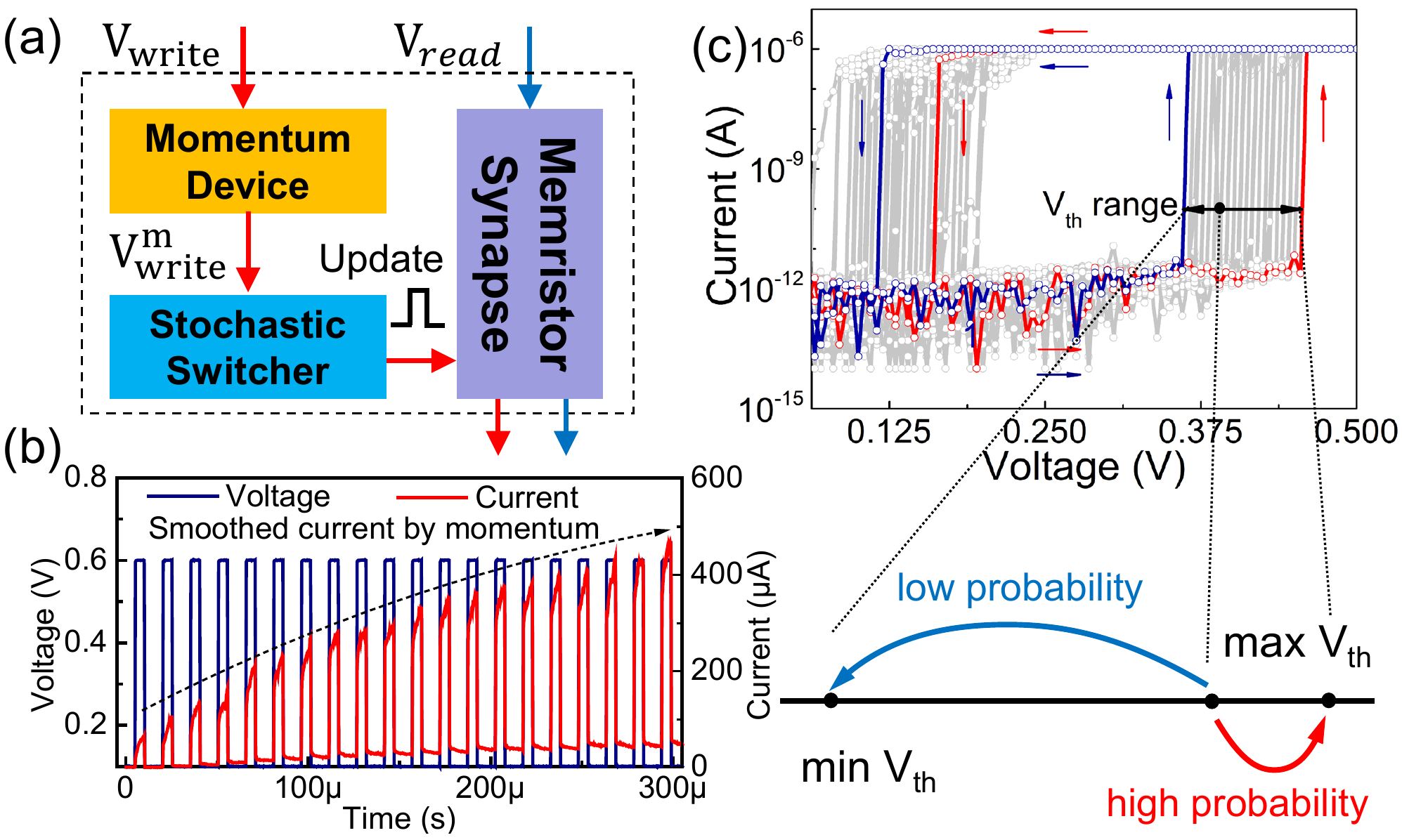}
\caption{(a) The scheme of inference and weight update of SSM. (b-c) Momentum and stochastic updating can be mimicked by diffusive memristors and volatile switching selectors, respectively.}
\label{fig15}
\end{figure}

The SSM scheme can be constructed not only by CMOS, such as field programmable gate array (FPGA) and the ASIC chip, but also has great potentials to be implemented by two types of emerging nano-electronic devices, as shown in Figure~\ref{fig15}. Recently, an exponentially conductance change characteristic is found in some memristors with short-term plasticity, such as the diffusive memristors, where the momentum can be mimicked. And the stochastic rounding behavior can be also accomplished by the volatile threshold switching selector devices, due to their variation of the threshold voltage.

\section{Conclusion}
In this work, we quantitatively analysis effects of various non-ideal characteristics in real memristor network on the performance of MLP and CNN. Meanwhile, based on the ablation studies, we design a novel SSM scheme to train the non-ideal memristor based neural network. Impressive classification accuracy is obtained with fewer epochs and programming pluses during the in-situ learning. The simulation results indicate that the SSM scheme promises a fast and low-power in-situ training solution for non-ideal memristor networks in-situ learning, which is an essential step to apply real memristor networks to practical applications.

\section*{Acknowledgments}
This work is supported by National Nature Science Foundation of China (Nos. 61327902 and 61836004), Suzhou-Tsinghua innovation leading program (No. 2016SZ0102) and Brain-Science Special Program of Beijing under Grant Z181100001518006

\bibliography{SSM}
\bibliographystyle{IEEEtran}

\end{document}